# Dynamic Web Service Composition based on Network Modeling with Statistical Analysis and Backtracking

M.SureshKumar[1] and P.Varalakshmi[2]

[1] Research Scholar, Anna University, Chennai,
`suresh.priya.kumar@gmail.com`

[2] Department of Information Technology, MIT Campus, Anna University, Chennai,
`varanip@gmail.com`

## ABSTRACT

*A Web service is a software system designed to support interoperable machine-to-machine interaction over a network. Web services provide a standard means of interoperating between different software applications, running on a variety of platforms and/or frameworks. One of the main advantages of the usage of web services is its ability to integrate with the other services through web service composition and realize the required functionality. This paper presents a new paradigm of dynamic web services composition using network analysis paired with backtracking. An algorithm called "Zeittafel" for the selection and scheduling of services that are to be composed is also presented. With the proposed system better percentage of job success rate is obtained compared to the existing methodology.*

## KEYWORDS

*Dynamic composition, PERT, backtracking, tour planner*

## 1. INTRODUCTION

Web services are considered as self-contained, self-describing, modular applications that can be published, located, and invoked across the Web. Nowadays, an increasing amount of companies and organizations only implement their core business and outsource other application services over Internet. Thus, the ability to efficiently and effectively select and integrate inter-organizational and heterogeneous services on the Web at runtime is an important step towards the development of the Web service applications. Web services can be engaged as in Figure 1. In particular, if no single Web service can satisfy the functionality required by the user, there should be a possibility to combine existing services together in order to fulfill the request. This trend has triggered a considerable number of research efforts on the composition of Web services both in academia and in industry. In the research related to Web services, several initiatives have been conducted with the intention to provide platforms and languages that will allow easy integration of heterogeneous systems. There are two methods for web services composition [10, 11, 12]. One is static web service composition and other is automated/dynamic web service composition. In static web service composition, composition is performed manually, that is each web service is executed one by one in order to achieve the desired goal/requirement. It is a time consuming task which requires a lot of effort. In automated web service composition, agents are used to select a web service that may be composed of multiple web services but from user's viewpoint, it is considered as a single service [13].

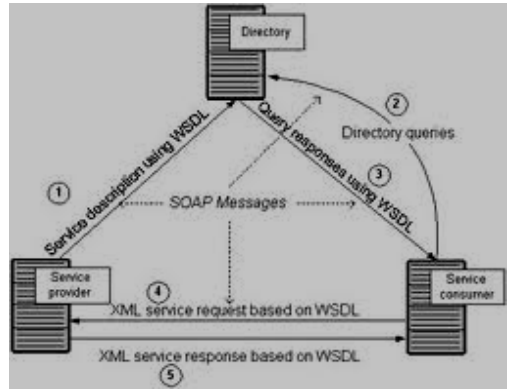

Figure 1: General Process of Engaging a Web Service

The integration of heterogeneous web services together to realize business functionality is called web service composition. In other words web service composition can be termed as an aggregation of individual web services to automate a particular task or a business process [2]. Through this paper we present tour planner software, "Web Safari", which uses web services composition by the proposed "Zeittafel" algorithm. In this software the various activities involved in the tour such as flight booking, hotel booking, scenic spot searching, dinner booking, etc are modeled as the individual nodes or activities of the Program Evaluation and Review Technique (PERT) network. There may be several individual service providers available for each activity.

Each of these is modeled as a web service. For example, various hotels may be present in the same locality offering same service. The best service is selected based on the availability of the service at that particular instant of time and various other constraints. This service selection is performed using the PERT method [3]. In case if no service is available in a particular category the processing may get into busy wait. In order to avoid this condition backtracking is employed [4]. This enables us to obtain an optimal solution and finally a set of services will be obtained that is the tour plan for the user. The reservation for these services can be done.

## 2. RELATED WORK

The QoS factors performance, reliability, availability and cost were dealt in the web service composition for e-business systems by Sayed Gholam and Hassan Tabatabaei [8]. This uses Web Service Modelling Ontology (WSMO). Web service composition using Web Service Databases (WSDB) is proposed by Farhan Hassan Khan, M.Younus Javed, and Saba Bashir [9]. It presented a framework in which multiple repositories and WSDBs have been introduced in order to make system more reliable and ensure data availability. Incheon Paik and Daisuke Maruyama [7] proposes a framework for automated web services composition through AI (Artificial Intelligence) planning technique by combining logical combination (HTN) and physical composition (CSP (Constraint Satisfaction Problem)). This paper discusses the real life problems on the web that is related to planning and scheduling. It provides task ordering to reach at desired goal. P.Sandhya and M.Lakshmi [3] deal with an algorithm for automatic quality driven web service called Opus deviser algorithm for business and task planners using PERT. Here infinite looping and exhaustive waiting problems arise. Gexin Li, Aixin Zu, Chengwen Wu, Zhengzhong Wang [4] deal with backtracking to arrive at optimal solutions for web service composition. This paper combines both PERT and backtracking where PERT is used for service evaluation and backtracking for service selection into the composition to obtain an optimal tour plan.

## 3. PERT AND BACKTRACKING

Network analysis is the general name given to certain specific techniques which can be used for the planning, management and control of projects. There are two widely applied network analysis models namely Critical Path Method (CPM) and Program Evaluation and Review Technique (PERT).

CPM was the trade-off between the cost of the project and its overall completion time (e.g. for certain activities it may be possible to decrease their completion times by spending more money - how does this affect the overall completion time of the project). PERT was used for the planning and control of the Polaris missile program and the emphasis was on completing the program in the shortest possible time. In addition PERT had the ability to cope with uncertain activity completion times (e.g. for a particular activity the most likely completion time is 4 weeks but it could be anywhere between 3 weeks and 8 weeks).

Due to the complex nature of most projects, it is very difficult to completely innate the delays and the cost overruns. However, with the appropriate management systems for planning, organizing, and controlling, it is possible to reduce them to a reasonable level. The problem is that the cost of implementing and executing such systems can exceed their benefits because of the large amount of monitoring and reporting that is required. The major purpose of PERT and CPM is to objectively identify these critical activities. Further, these techniques can tell us how close the remaining activities are to becoming critical. (This available delay is called slack or float.).

PERT and CPM are very similar in their approach however, two distinctions are usually made. The first relates to the way in which activity duration are estimated. In PERT, three estimates are used to form a weighted average of the expected completion time, based on a probability distribution of completion times. Therefore, PERT is considered a probabilistic tool. In CPM, there is only one estimate of duration; that is, CPM is a deterministic tool. The second difference is that CPM allows an explicit estimate of costs in addition to time. Thus, while PERT is basically a tool for planning and control of time, CPM can be used to control both the time and the cost of the project. Extensions of both PERT and CPM allow the user to manage other resources in addition to time and money, to trade off resources, to analyze different types of schedules, and to balance the use of resources.

Program Evaluation and Review Technique (PERT) [5] is a statistical tool used in project management that is designed to analyze and represent the tasks involved in the given project. One key element to PERT's application is that four estimates are required because of the element of uncertainty and to provide time frames for the PERT network. It is probabilistic in nature and involves the calculation of

*1. Optimistic time (O):* The minimum possible time required to accomplish a task, assuming everything proceeds better than is normally expected.

*2. Pessimistic time (P):* The maximum possible time required to accomplish a task, assuming everything goes wrong (but excluding major catastrophes).

*3. Most likely time (M):* The best estimate of the time required to accomplish a task, assuming everything proceeds as normal.

*4. Expected time ($T_E$):* The best estimate of the time required to accomplish a task, accounting for the fact that things don't always proceed as normal (the implication being that the expected time is the average time the task would require if the task were repeated on a number of occasions over an extended period of time). It can be calculated using the formula specified in equation (1).

$$T_E = (O + 4M + P) / 6 \quad \text{-------------------} (1)$$

Two other elements comprise the PERT network: the path, or critical path, and slack time. The critical path is a combination of events and activities that will necessitate the greatest expected completion time. Slack time is defined as the difference between the total expected activity time for the project and the actual time for the entire project. Slack time is the spare time experienced in the PERT network.

Backtracking [6] is a general algorithm or a tool to arrive to a solution for constraint satisfaction problems. A classic example for this is the N-queens problem. Backtracking is also used to optimize the solution from a given set of partial candidate solutions.

## 4. PROPOSED WORK

The proposed work can be split into 2 steps.

1. Selection of services
2. Composition of services

*Step 1: Selection of services*

As stated before several services are available in each category. Only one of them must be selected from each of the categories. The primary constraint that we seek is the availability of the service. In the case of tour planning, say once a flight is completely booked the service can be blocked in the registry. Thus if the reservation is not available then the service is also not available. Thus first and foremost the availability factor of the Quality of Service constraints is checked. Once the service is available it is counted as an option for the future processing. Secondly with respect to the tour planner, the constraints of the user such as cost, number of people, etc are applied and some services are filtered out.

All the available services are constructed as a set. Some of the services may need to follow a particular sequence while some may not. For the services that follow a predefined order the normal Opus deviser algorithm may be followed. But for the activities where the sequence is not important, for example sight seeing, then the modified Opus deviser algorithm is followed. This makes use of the PERT method. In some cases no service in a category may be available. In such situations the backtracking is employed in order to change the order of activities or to intimate the user about the failure. This rescues the system from entering into the busy wait condition.Finally a single solution or a set of solutions can be obtained. If multiple solutions are available the choice is left to the user.

*Step 2: Composition of services*

The selected set of services is composed together and orchestrated using BPEL engine. The booking for the required spots are done and the tour plan is finalized.

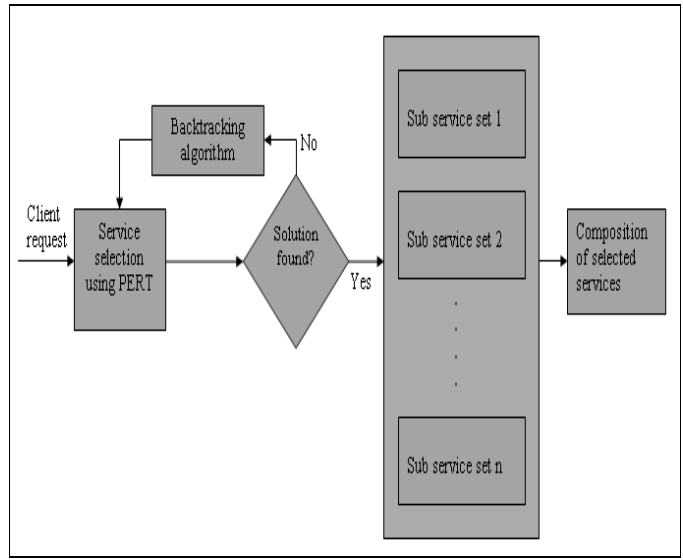

Figure 2: Architecture of proposed system

## 5. ALGORITHM

Our algorithm named **"Zeittafel"** is a PERT model based algorithm using backtracking to handle exceptions.

*Input:*
- Set of web services

$$S = \begin{bmatrix} S_{11} & \cdots & S_{1n} \\ \vdots & & \vdots \\ S_{m1} & \cdots & S_{mn} \end{bmatrix}$$

Where m= number of web services in each category
n= number of web service categories
- Goal deadline (g)
- Fixed sequence of categories (where applicable) FC= [ $fc_1$ $fc_2$ ….. $fc_p$] where p<=n
- Non-fixed sequence of categories (generated randomly)
NC= [ $nc_1$ $nc_2$ ….. $nc_q$] where q<=n and p+q=n.

*Output:* Composed web services

*Algorithm:*

1. Begin

2. Build a matrix of web services W such that there is at least one service that is available from each category. Therefore W ⊆ S

3. Generate all partial candidate combinations from the matrix W such that there is at least one service from each category following the given sequence of categories FC wherever applicable along with the randomly generated NC. The solution must be performed within the goal deadline.

4. If successful combination(s) is(are) generated then
   Go to step 5.

   Else

   a. Backtrack to change the order of the non-sequential category NC. Replace NC with changed NC by changing the categories in a circular fashion until the circle is completed. Go to step 2.

   b. If no solution can be found then negotiate with client for the withdrawal of certain non available sub categories of the service. Go to step 2.

5. For each combination PERT network analysis model is applied.

   a. Let optimistic time as O, most likely time as M, pessimistic time as P for each activity of the PERT model.
   b. Compute expected time T for each activity as T= (O+4M+P)/6.
   c. Create a directed graphical representation of each combination with each event as a node and each activity as an arrow. Compute the float time for each activity and the critical path.
   d. Calculate the variance for each activity in all combinations as vc= $((P-O)/6)^2$. Thus the set of variances become V= {$vc_1$, $vc_2$, ....., $vc_n$}
   e. Compute the variance of critical path of each combination as CV={ $cv_1$, $cv_2$,...., $cv_n$} where $cv_c$= sum of variances of each activity in the critical path of the $c^{th}$ combination.

   $$cv_c = \sum_{1}^{n} vc_c \text{ ------------------- } (2)$$

   f. Compute standard deviation for each combination as
   $$SD_c = \sqrt{cv_c} \text{ -------------------- } (3)$$
   g. In order to achieve the task within the goal deadline g, assign X=g where X is the time under consideration.
   h. Apply normal distribution of probability with normal variate Z={$z_1,z_2,....,z_n$}. $z_i$= (X-$x_i$)/$SD_i$ where $x_i$ is the critical path time of the combination.
   i. Find the probability of completion of the project with the given sequence from the normal table and construct the probability set A={$a_1,a_2,....,a_n$}

6. Select a combination such that the probability of completion is high. If probability remains same for more combinations, the one with low critical path is selected. User intervention may also be preferred. This optimizes the solution.

7. The services that are within the selected combination are composed.

8. Stop

# 6. IMPLEMENTATION AND RESULTS

This project has been implemented in J2EE framework. J2EE web services were developed with Axis 2 engine. Their respective WSDL files were generated and thus they were added to the registry. Another web service that implements the Zeittafel algorithm is developed.

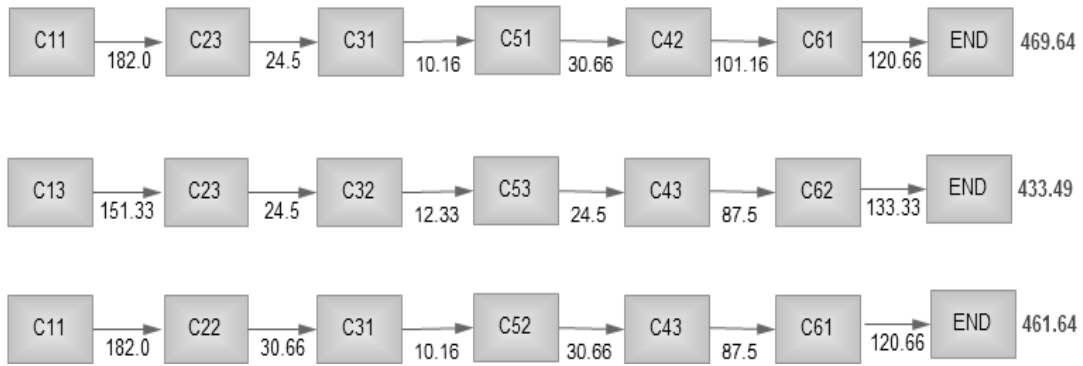

Figure 3: Examples of calculating task completion time using PERT. The data in red shows the total expected time taken to complete the activity

This involves the PERT and backtracking and it selects the web services for composition. The example PERT representation is depicted in Figure 3. These are composed using BPEL engine and are orchestrated together.

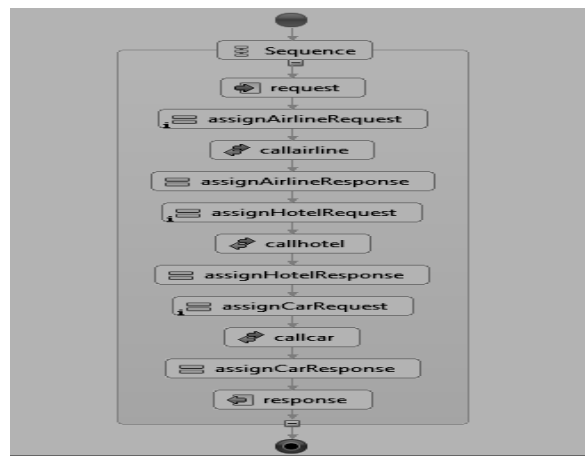

Figure 4: BPEL implementation

Business Process Execution Language (BPEL), short for Web Services Business Process Execution Language (WS-BPEL) is a standard executable language for specifying actions within business processes with web services. Processes in BPEL export and import information by using web service interfaces exclusively. WS-BPEL 2.0 is an XML-based language for defining business processes that orchestrate web services and this is used in the project. The implementation uses Riftsaw Open Source JBOSS BPEL engine.

The UI part of the algorithm gets the necessary input from the user such as deadline, etc and invokes the algorithm.

We voluntarily introduce situations in the system that lead to the invocation of backtracking.
The deadline g is given as 450 minutes.

With respect to the "Web Safari" project the implementation is described as below. There are 6 categories of web service providers such as flight, taxi, hotel, tourist spot1, tourist spot2 and tourist spot3. Each of the categories has 3 web services. They are named C1 through C6 respectively. The first 3 categories are fixed and the rest of them are non-fixed.

i.e  FC={ C1, C2, C3}
     NC= {C4, C5, C6}

The constraints for backtracking is introduced such that C4 is not available at the required point of time but if the sequence NC= {C5, C4, C6} the output is obtained. The remaining combinations do not meet the constraints. The following table (Table 1) describes the time taken in minutes to complete each task in each category by the web services.

Table 1: Time taken to complete each task (in mins)

| Cateogory | WS1 | WS2 | WS3 |
|---|---|---|---|
| C1 | 180 | 210 | 150 |
| C2 | 20 | 30 | 25 |
| C3 | 10 | 12 | 15 |
| C4 | 90 | 100 | 85 |
| C5 | 30 | 30 | 25 |
| C6 | 120 | 135 | 125 |

For example, to reach the given place through flight represented by WS1 in C1 (say Air India) 180 minutes is required. With these given values various combinations are generated and the program is executed. This also involves the random non-sequential category generation once it is found the C4 is not available. To demonstrate the working a few examples are shown in figure 3. The probability of task completion is also calculated and a normal distribution curve for the combination is drawn. A bell-shaped normal distribution curve as shown in figure 5 is obtained. The region in dark line denotes the probability of completion.

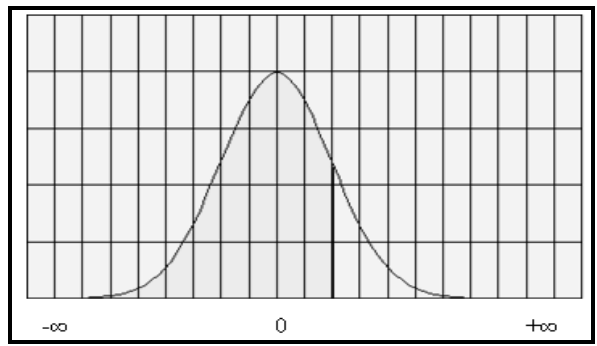

Figure 5: Normal distribution curve

Thus the combination with least completion time is selected and this is composed together and the tour plan is presented to the user.

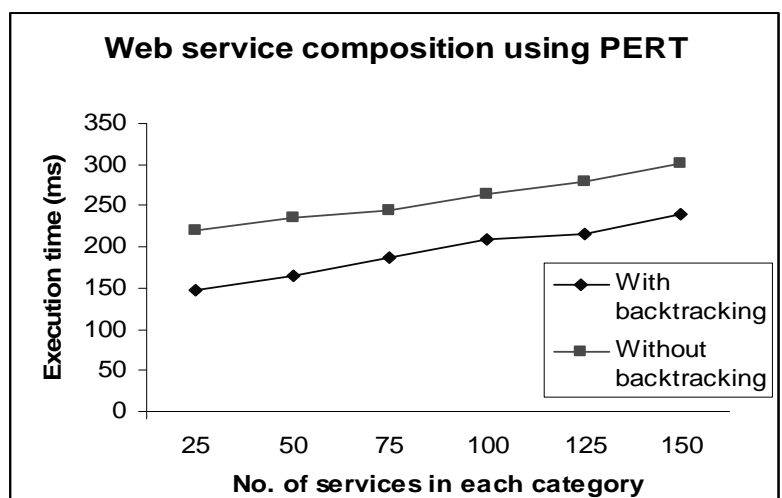

Figure 6: Efficiency of our algorithm

The same data is input to the program without backtracking. In that case the execution time of the system with backtracking is efficient than that without backtracking. Figure 6 represents the difference between the systems with and without backtracking. The system constructed has constraints where backtracking logic has to be invoked. Though the execution time will be increased due to the invocation of backtracking logic the overall process completion time of the system will be less than the system without backtracking due to the fact that the system enters into an exhaustive waiting state.

## 7. CONCLUSION

The uncertain nature of the completion of activities is taken into account. Highly optimized dynamic web service composition with provisions for reliability in the form of backtracking is introduced. QoS of availability is dealt here with respect to the blocking of services that are not available for the given point of time along with the filtration of them in the matrix.

**M.SureshKumar** received the B.E degree in Computer Science & Engineering from Madras University, TamilNadu, India and the M.E Degree in Computer Science & Engineering from Sathyabama University, India. He is working as an Assistant Professor in Information Technology Department in Sri Sai Ram Engineering College-Chennai. 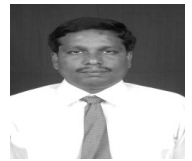

**Dr.P.Varalakshmi** obtained Ph.D. from the Faculty of Information and Communication Engineering at Anna University. She is working as a Assistant Professor in the Department of Information Technology at Madras Institute of Technology, Anna University. She has published many research papers in international and national journals and conference proceedings with very high citation index of about 25 papers so far. 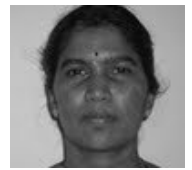